\documentclass[useAMS,usenatbib]{mnras}
\usepackage{graphicx}
\usepackage{amsmath}
\usepackage{amssymb}
\usepackage{color}

\def \etal {et~al.~}
\def \msun {\ifmmode \rm M_{\odot} \else $\rm M_{\odot}$ \fi}
\newcommand{\Mpch}{{\ifmmode{h^{-1}{\rm Mpc}}\else{$h^{-1}$Mpc}\fi}}
\newcommand{\kms}{{\ifmmode{ {\rm km\,s^{-1}} }\else{ ${\rm km\,s^{-1}}$ }\fi}}

\newcommand{\Vmax}{\ifmmode{V_{\rm max}^{\rm DM}}\else{$V_{\rm max}^{\rm DM}$}\fi}
\def \rhocrit {\ifmmode \rho_{\rm crit} \else $\rho_{\rm crit}$ \fi}

\title[Einasto parameters of $\Lambda$CDM halos]{A deeper look into the structure of $\Lambda$CDM haloes: correlations between halo parameters from Einasto fits}
\author[Udrescu \etal]{Silviu M. Udrescu$^1$\thanks{E-mail: smu226@nyu.edu},  Aaron A. Dutton$^{1}$\thanks{E-mail: dutton@nyu.edu}, Andrea V. Macci\`{o}$^{1,2}$\thanks{E-mail: maccio@nyu.edu}, Tobias Buck$^2$\\ 
 $^{1}$New York University Abu Dhabi, PO Box 129188, Saadiyat Island, Abu Dhabi, United Arab Emirates\\
 $^{2}$Max Planck Institute f\"{u}r Astronomie, K\"{o}nigstuhl 17, 69117 Heidelberg, Germany\\
}
\begin{document}

\maketitle

\label{firstpage}

\begin{abstract}
  We used high resolution dark matter only cosmological simulations to
  investigate the structural properties of Lambda Cold Dark Matter
  ($\Lambda$CDM) haloes over cosmic time. The haloes in our study
  range in mass from $\sim 10^{10}$ to $\sim 10^{12} \msun$, and are
  resolved with $10^5$ to $10^7$ particles. We fit the spherically
  averaged density profiles of DM haloes with the three parameter
  Einasto function.  For our sample of haloes, the Einasto shape
  parameter, $\alpha$, is uncorrelated with the concentration, $c$, at
  fixed halo mass, and at all redshifts. Previous reports of an
  anti-correlation are traced to fitting degeneracies, which our fits
  are less sensitive to due to our higher spatial resolution. However,
  for individual haloes the evolution in $\alpha$ and $c$ is
  anti-correlated: at redshift $z=7$, $\alpha \simeq 0.4$ and
  decreases with time, while $c\simeq 3$ and increases with time. The
  evolution in structure is primarily due to accretion of mass at
  larger radii. We suggest that $\alpha$ traces the evolutionary state
  of the halo, with dynamically young haloes having high $\alpha$
  (closer to a top-hat: $\alpha^{-1}=0$), and dynamically relaxed
  haloes having low $\alpha$ (closer to isothermal: $\alpha=0$). Such
  an evolutionary dependence reconciles the increase of $\alpha$ vs
  peak height, $\nu$, with the dependence on the slope of the power
  spectrum of initial density fluctuations found by previous studies.
\end{abstract}


\section{Introduction}

Accurately predicting structural properties of cold dark matter (CDM)
haloes is one of the fundamental goals of modern
cosmology. Predictions can be used to test CDM using rotation curves
\citep[e.g.][]{Moore94,deBlok08}, and as inputs to  analytic and
semi-analytic galaxy formation models
\citep[e.g.][]{Dutton07,Dutton09}.  Due to the non-linear nature of
the problem, the standard approach is direct computation, starting
from primordial density fluctuations, and evolving the system under
the influence of gravity. 

Early simulations found a central cusp with density profile
$\rho\propto r^{-1}$ \citep{Dubinski91}. This is similar to the
Hernquist profile \citep{Hernquist90} commonly used to model the
stellar distribution in elliptical galaxies. Over large radii DM
haloes are more extended, and are better described by a two parameter
NFW profile with inner slope $-1$, and outer slope $-3$
\citep{Navarro96,Navarro97}.  A common parameterization is to use the
halo mass, $M$,  and the concentration parameter, $c$, which is the
ratio between the virial radius and the characteristic scale radius,
and is hence dimensionless. Concentration and mass are highly
correlated \citep{Bullock01,Maccio07}, so that to first order the
structure of CDM haloes depend just on the halo mass.

The dependence of halo concentration on mass, redshift, and
cosmological parameters has been the subject of extensive computational study.
Recent work has shown that the three parameter Einasto profile \citep{Einasto65}
\begin{equation}
\label{eq:einasto}
  \frac{\rho_{\rm EIN}}{\rho_{-2}} = \exp \left\{ - \frac{2}{\alpha} [ (r/r_{-2})^{\alpha} -1 ] \right\}, 
\end{equation}
provides a better description of CDM density profiles
than either NFW \citep{Gao08,Dutton14,Navarro2010,Klypin16} or a
three parameter generalized NFW \citep{Klypin16}.

The extra parameter relative to an NFW profile, $\alpha$, controls the
deviation from an isothermal distribution (i.e., $\rho = r^{-2}$):
$\alpha=0$ corresponds to an isothermal profile, while $\alpha = 2$
corresponds to a Gaussian, and $\alpha=\infty$ corresponds to a
uniform density sphere. Thus haloes with larger $\alpha$ have less
mass at small and large radii relative to haloes with lower $\alpha$.

The shape parameter is an increasing function of halo mass and
redshift.  Strikingly the dependence with redshift is completely
removed when the halo mass is replaced with the dimensionless peak
height of the halo \citep{Gao08,Prada12,Dutton14,Klypin16}. This
peak height, $\nu$, is defined by
\begin{equation}
\label{eq:peakheight}  
  \nu(M,z) = \delta_{\rm crit}(z)/\sigma(M,z),
\end{equation}
where $\delta_{\rm crit}(z)$ is the linear density threshold for
collapse at redshift $z$, and $\sigma(M, z)$ is the rms linear density
fluctuation at $z$ within spheres of mean enclosed mass, $M$. The
parameter $\nu(M, z)$ is related to the abundance of objects of mass
$M$ at redshift $z$. The characteristic mass $M_*(z)$ of the halo mass
distribution at redshift $z$ is defined through $\nu(M_*,z)=1$.
Thus, haloes with $\nu=1$
are typical haloes and have $\alpha\simeq 0.16$ at all redshifts,
while haloes with $\nu=3$ are rare $3\sigma$ high fluctuations, and have
$\alpha\simeq 0.25$.

\citet{Cen14} outlined how the S\'ersic profile (which has the same
functional form as the Einasto profile, but applied to a 2D density)
could naturally arise out of a Gaussian random field. Based on this,
\citet{Nipoti15} ran cosmological simulations and found a relation
between the shape of the dark matter density profile with the index of
the power spectrum of initial density fluctuations. Haloes formed in a
universe with flat spectra (e.g., $n=0$) have the mass in fewer larger
clumps, while haloes formed from steep spectra (e.g., $n=-3$) have a
larger amount of the mass in small clumps. Flat spectra resulted in
lower $\alpha\simeq 0.1$ vs $\alpha\simeq 0.3$ for steep spectra.
Qualitatively similar results have been  reported by  \citet{Ludlow17}
using power-law initial power spectra (lower-$n$ results in higher
$\alpha$ at fixed peak height). 

Based on this insight, since higher mass haloes form on scales where
$n$ is larger, one would nominally predict that high mass haloes
should have slightly lower $\alpha$.  However, exactly the opposite is
seen in cosmological simulations: higher mass haloes have
significantly higher $\alpha$ \citep{Gao08,Prada12,Dutton14}.

There are structural differences between haloes of different masses:
more massive haloes are, on average, less concentrated by $\simeq 0.1$
dex per decade in halo mass. If less concentrated haloes have higher
$\alpha$ this would help explain why we don't see $\alpha$ decrease
with increasing halo mass or peak height.  Indeed, \citep{Ludlow13}
reported a negative correlation: haloes offset to high $c$ from the
concentration mass relation had lower $\alpha$ while haloes offset to
low $c$ had higher $\alpha$.  This correlation seems to be consistent
with the fact that  young haloes (i.e., high peak heights, $\nu$) are
known to have low $c$ and high $\alpha$ \citep{Dutton14}.  At issue is
the relatively low resolution of the simulated haloes and thus the
sensitivity to a fitting degeneracy between $c$ and $\alpha$.

In this paper we investigate further the correlations between Einasto
shape parameter and concentration and the evolution of these
parameters with time. We use zoom-in simulations in order to obtain
much higher spatial resolution than is possible with cosmological
volumes. We use two sets of simulations: Milky Way mass haloes with
$\sim 10$ million particles per halo, and dark matter only counter
parts to the NIHAO suite with $\sim 1$ million particles per halo. 

This paper is organized as follows: in Section \ref{sec:sims} we will
briefly summarize the properties of the simulations, in Section
\ref{sec:res} we present the scaling relations, while in Section
\ref{sec:evolution} we present the evolution of the most massive halo
in each  simulation. In Section \ref{sec:disc} we discuss our results
and then conclude with a summary in Section \ref{sec:conc}.

\begin{figure}
\includegraphics[width=0.5\textwidth]{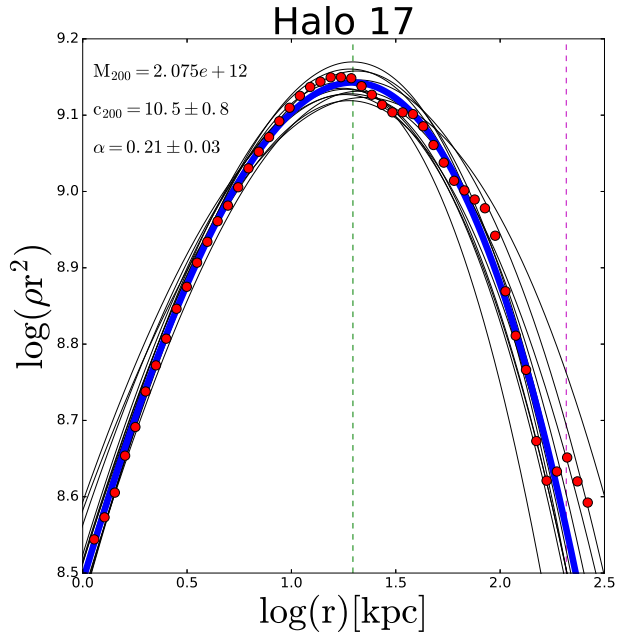}
\caption{Example of an Einasto fit to a $10^{12}$ $\msun$ halo (thick
  blue line). The red points show the density profile computed in 50
  logarithmically spaced bins from three times the softening length to
  twice the virial radius.  The green dashed line marks the position of the
  scale radius, $r_{-2}$ and the virial radius, $R_{200}$ is marked by
  the magenta dashed line. The grey lines represents 12 samples from the
  chain, obtained during the fitting procedure.}

\label{fig:example}
\end{figure}

\begin{figure*}
  \centerline{
    \includegraphics[width=0.98\textwidth]{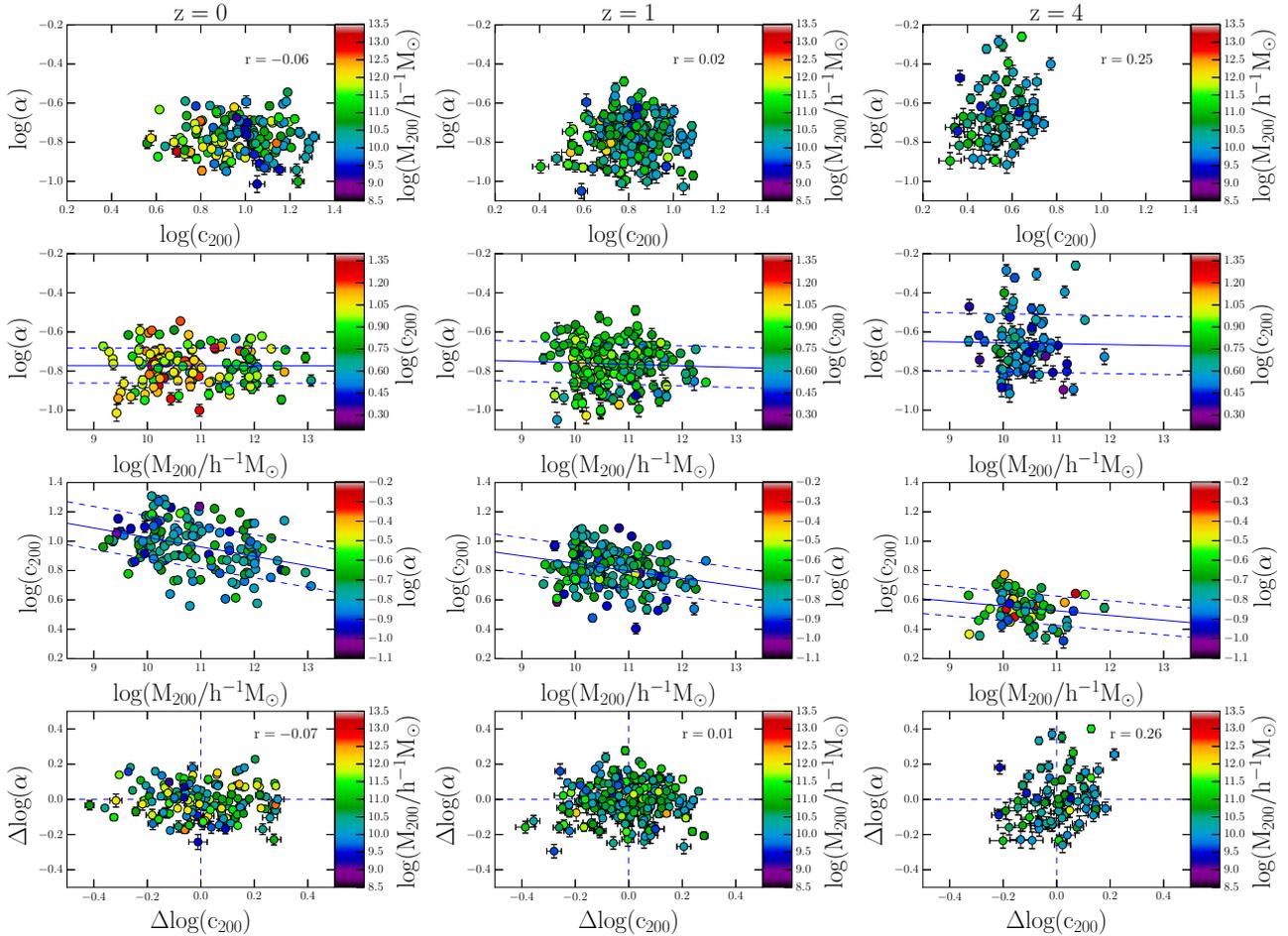}
 }
\caption{Relations between concentration, $c$, Einasto shape
  parameter, $\alpha$ and halo mass, $M_{200}$, at redshifts $z=0$
  (left), $z=1$ (middle), and $z=4$ (right). We include only relaxed,
  non-polluted haloes with at least $10^5$ particles.  Points are
  colour coded by the 3rd parameter (right colour-bar axis). The bottom
  row shows the residuals of the c vs $M_{200}$ and $\alpha$ vs
  $M_{200}$ relations at each redshift.  These plots show there  is no
  significant correlation between $c$ and $\alpha$.}
\label{fig:cam}
\end{figure*}

\section{Simulations}
\label{sec:sims}

We primarily use simulations from two projects.  The
\citet{Buck15, Buck16} study of 21 haloes of present day mass $\sim
10^{12}\msun$, and the dark matter only counterparts of the NIHAO
\citep{Wang15} study of 91 haloes of present day masses from $\sim
10^{10}$ to $\sim 10^{12}\msun$.  Each halo is simulated using the
``zoom-in'' technique, so there is one main halo per simulation, plus
several field haloes within the zoom-in region.  The main halo
contains $\sim 1$ million particles in NIHAO, and $\sim 10$ million
particles in Buck \etal

Haloes were identified using the MPI+OpenMP hybrid halo finder
\texttt{AHF}\footnote{http://popia.ft.uam.es/AMIGA} \citep{Gill04,
  Knollmann09}. \texttt{AHF} locates local over-densities in an
adaptively smoothed density field as prospective halo centers. The
virial masses of the haloes are defined as the masses within a sphere
whose average density is 200 times the cosmic critical matter density,
$\rhocrit=3H_0^2/8\pi G$.  The virial mass, size and circular velocity
of the hydro simulations are denoted: $M_{200}, R_{200}, V_{200}$.

\citet{Power03} gave a number of convergence
criteria for N-body simulations. For our purposes the strictest
requirement is the relaxation time (their equation 20). For resolving
the scale radius, this can be approximated with $N_{\rm min} =
16000(c_{200} /10)^2$ \citep{Dutton14}. 
In order to constrain the $\alpha$ parameter, we wish to resolve
scales significantly smaller than $r_{-2}$. Thus we impose a minimum
particle number of $N_{\rm min}=10^5$. For the Buck \etal haloes this
results in a minimum halo mass of $\sim 10^{10}\msun$.  We require
that the mass fraction of low-resolution (i.e., polluting) particles
be $f_{\rm poll} < 10^{-3}$. 

In order to filter out merging haloes (which have ill defined
structural parameters) we measure two parameters related to relaxation
state of the halo following \citep{Maccio07}, $x_{\rm off}$ which is
the distance between the most bound particle and the center of mass,
in units of the virial radius, and $\rho_{\rm rms}$ which is the mean
deviation ($\log_{10}\rho_{\rm dm}$) between data and fit. For our
relaxed halo criteria we adopt $x_{\rm off} < 0.1$, and $\rho_{\rm
  rms} < 0.1$. We also put cuts on the error in log space of
$c_{200}$ and $\alpha$, such that
$e_{\log c} < 0.05$ and $e_{\log \alpha} < 0.05$. In the end, out of
the initially selected haloes, we kept for analysis $74 \%$ at
redshift $z=0$, $56 \%$ at $z=1$ and $41\%$  and $z=4$.

\begin{figure*}
\includegraphics[width=0.49\textwidth]{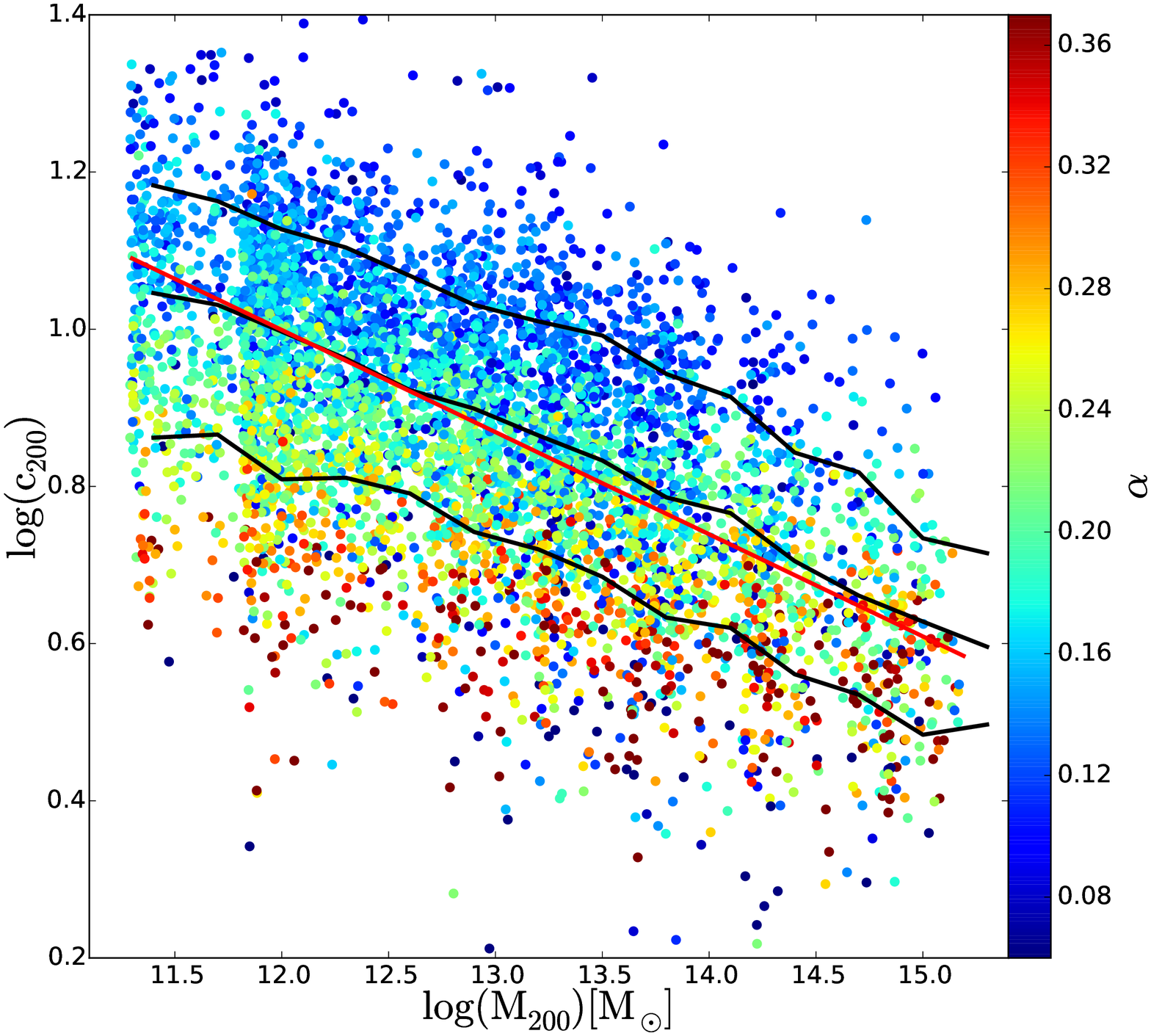}
\includegraphics[width=0.49\textwidth]{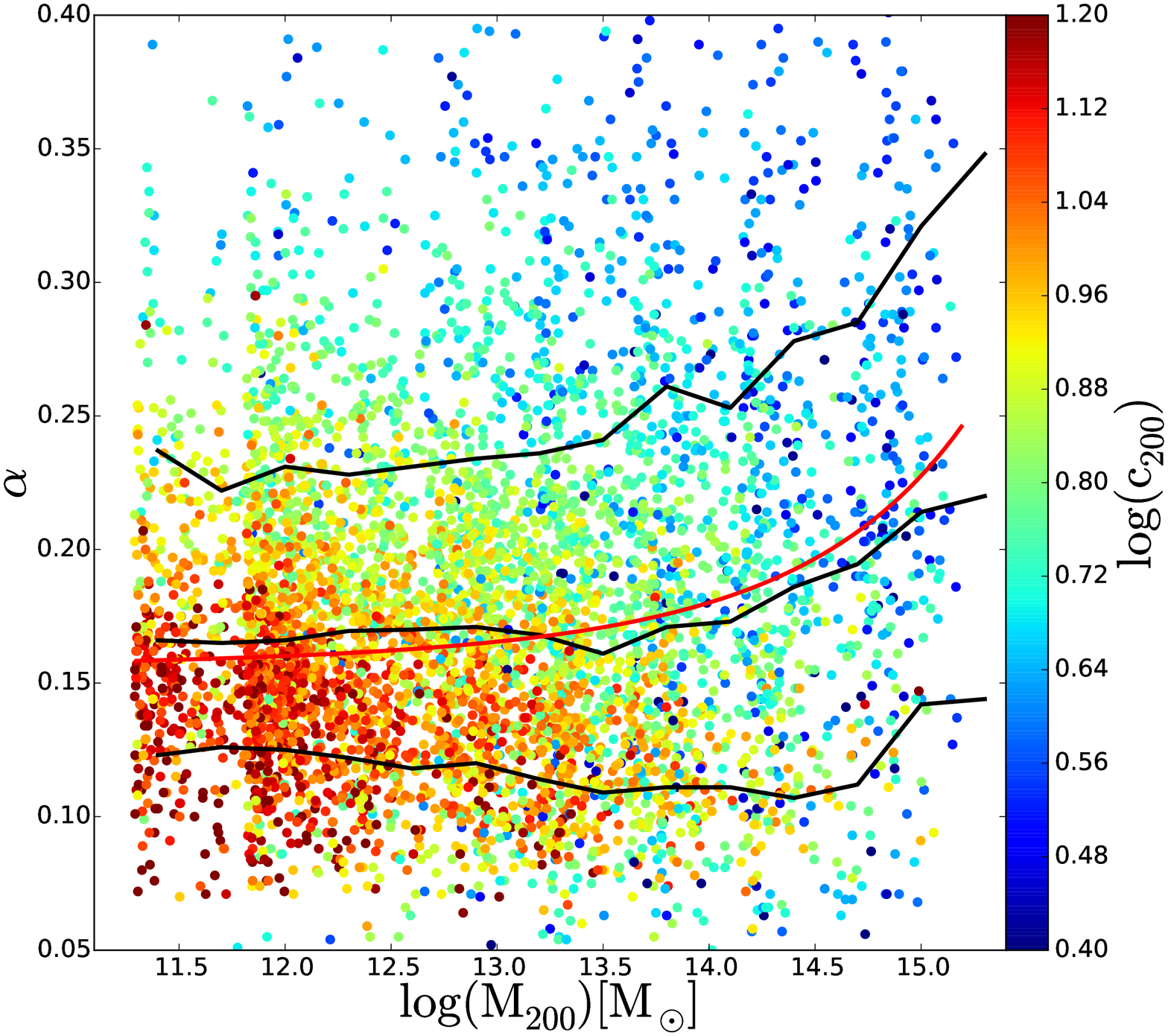}
\caption{Concentration vs halo mass (left) and Einasto shape parameter
  vs halo mass (right) from low resolution ($N \sim 10000$)
  cosmological volume simulations \citep{Dutton14}. The points are
  colour coded according to $\alpha$ (left) and $c$ (right), showing a
  clear anti-correlation between $\alpha$ and $c$ at fixed halo
  mass. The black lines show the 16th, 50th, and 84th percentiles in
  bins of halo mass. The red lines show the relations from
  \citet{Dutton14} (top) and \citet{Gao08} (bottom).}
\label{fig:cam_planck}
\end{figure*}

\subsection{Fitting procedure}
Spherically averaged density profiles were computed for all haloes
that pass the described filters. For each halo, the space between 3
times the softening length and 1.5 $R_{200}$ was divided into 50
shells, equally spaced in log radius. In order to obtain the density
profile, for each shell the distance from the center was calculated as
the average of the inner and the outer radius of the shell, while the
density of the shell was computed using the mass of all the particle
inside the shell and its volume.

We fit each halo with the three parameter Einasto function
(Eq.~\ref{eq:einasto}) which is specified by the shape parameter,
$\alpha$, scale radius, $r_{-2}$, and normalization, $\rho_{-2}$.
Since fitting the Einasto function is non-linear, we use the Monte
Carlo Markov Chain (MCMC) code {\sc emcee} \citep{emcee} to properly
sample the posterior distribution.  Initial best fit parameters are
obtained using the iMinuit package. The Priors for {\sc emcee} are
uniform from 0.1 to 10 times the initial best fit.  The best fit value
from {\sc emcee} is taken as median of the distribution, providing
better fits than the ones obtained form the iMinuit results. The
errors on the parameters for each fit are taken to be the $16^{th}$
and the $84^{th}$ percentile of the posterior distribution for the
given parameter.

An example fit is shown in Fig.~\ref{fig:example}. The blue line is
the obtained fit to the data, while the grey lines show 12 samples
obtained from the chain, during the fitting procedure. The green and
magenta dashed lines mark the scale radius, $r_{-2}$ and virial
$R_{200}$, respectively. Note that the concentration is computed using the
virial radius from the particle data (rather than from the fitted
profile), so is equivalent to the scale radius from the fit, but has
an advantage of being dimensionless.

\section{Scaling Relations}
\label{sec:res}

Fig.~\ref{fig:cam} shows the relations between concentration, $c$,
shape parameter, $\alpha$, and halo mass, $M_{200}$, at redshifts
$z=0,1$, and 4. The $c$ vs mass relation and $\alpha$ vs mass
relation have slopes, normalizations and evolutions in line with
previous studies \citep[e.g.,][]{Dutton14}.

We test for a correlation between $\alpha$ and $c$ using two
plots. The upper panels show a direct comparison, while the lower
panels show the residuals of the $c$ vs $M_{200}$ and $\alpha$ vs
$M_{200}$ relations. We find no significant correlation between
$\alpha$ and $c$ using either method at all redshifts (the correlation
coefficients are shown in the top right corner of each panel). We
have verified this lack of a correlation also exists when we include
unrelaxed haloes, and when we use less strict relaxation criteria.
Our result should be compared with Fig. 2 in \citep{Ludlow13}, which shows a
strong anti-correlation between $\alpha$ and $c$.

  In order to reproduce the result from \citet{Ludlow13} we use the
cosmological simulations from \citet{Dutton14}. These cover a similar
halo mass range with similar resolution as those used by
\citet{Ludlow13}. As with \citet{Ludlow13} we only consider haloes
containing at least $N=5000$ particles.
Fig.~\ref{fig:cam_planck} shows
the $c$ vs halo mass (upper) and $\alpha$ vs halo mass (lower)
relations at redshift $z=0$, colour coded by $\alpha$ and $c$,
respectively. These show very similar trends as reported by
\citet{Ludlow13}, namely, at a given halo mass, higher $c$ haloes have
on average lower $\alpha$.

\begin{figure}
  \includegraphics[width=0.42\textwidth]{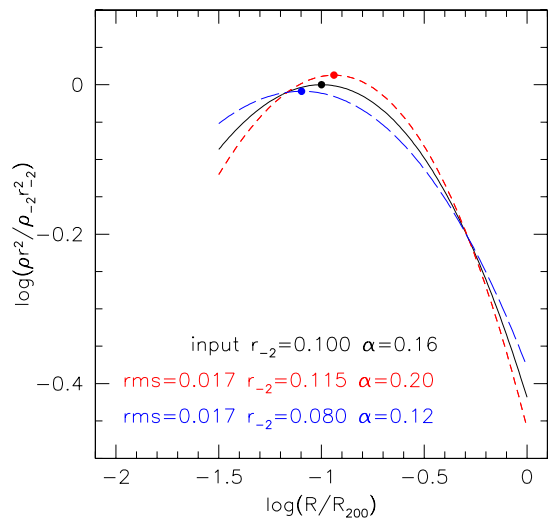}
\caption{ Illustration of the degeneracy between scale radius,
  $r_{-2}$, and shape parameter, $\alpha$, for the Einasto
  profile. The black line shows a profile with $r_{-2}=0.1$ and
  $\alpha=0.16$ and the black circle shows the location of the scale
  radius. The red and blue lines show profiles with the same
  $rms=0.017$. The red line has 25\% higher $\alpha$ and 11\% higher $r_{-2}$
  while the blue line has 25\% lower $\alpha$ and 20\% lower $r_{-2}$.}
\label{fig:degeneracy1}
\end{figure}

\begin{figure}
\includegraphics[width=0.48\textwidth]{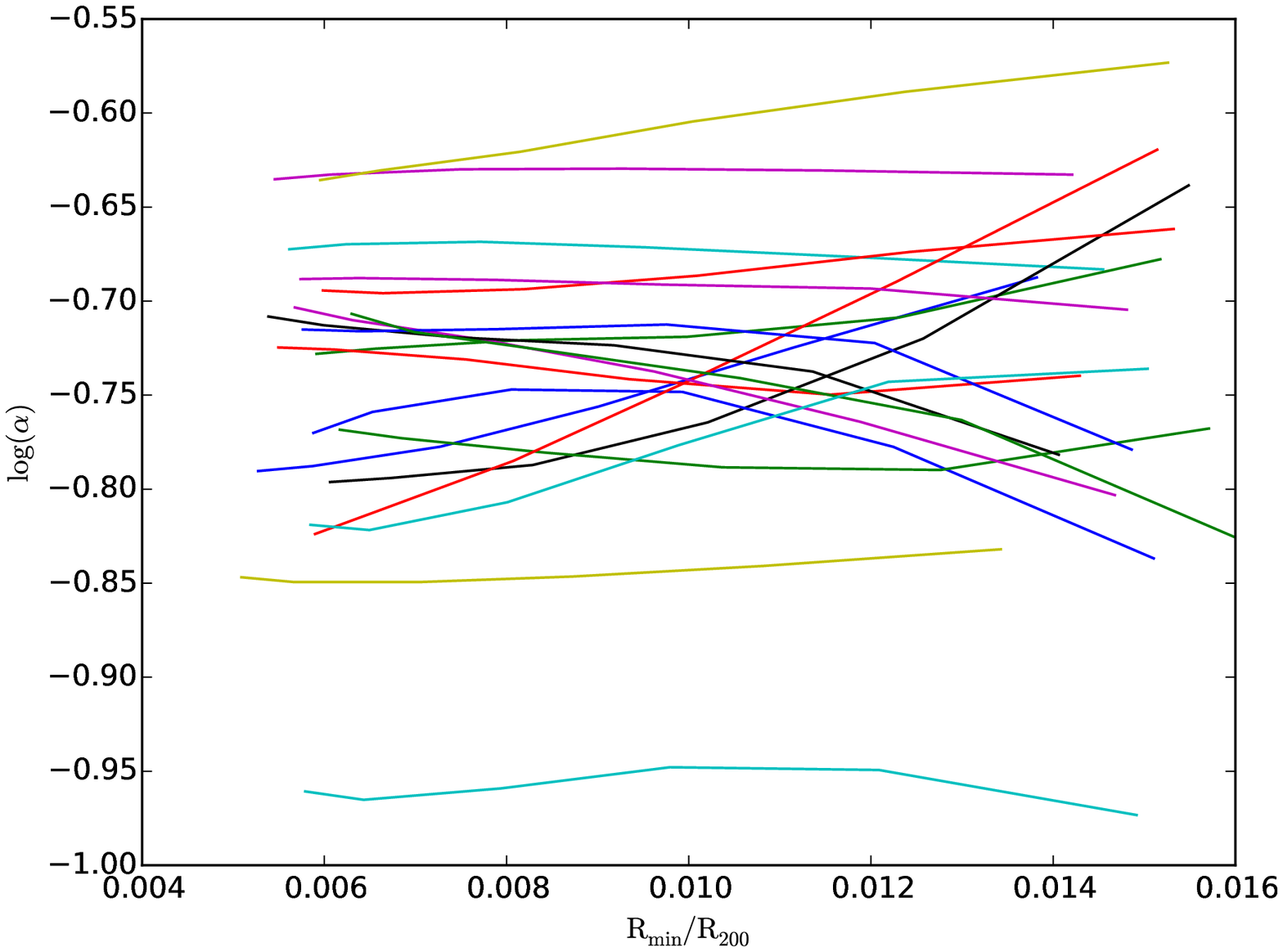}
\includegraphics[width=0.48\textwidth]{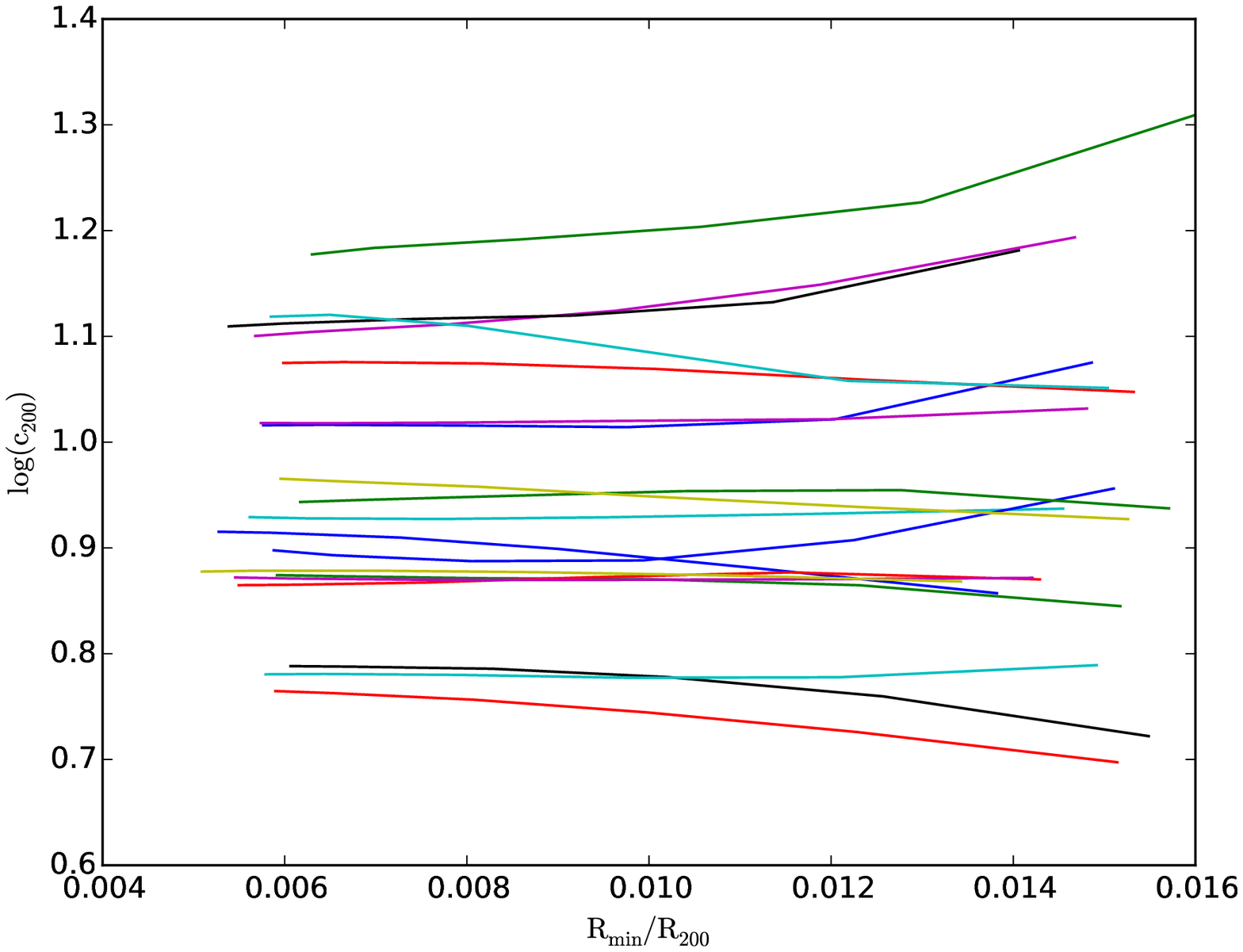}
\caption{ Impact of minimum radius on the fitted $\alpha$ (upper panel)
  and $c$ (lower panel) for the Buck haloes.  Each colour represents a
  different halo.}
\label{fig:rmin}
\end{figure}

\subsection{Impact of resolution on covariance between $c$ and $\alpha$}

An important difference between the haloes used in Figs.~\ref{fig:cam}
\& \ref{fig:cam_planck} is the number of particles. The number of
particles is directly related to the smallest scales that the
simulation can resolve \citep{Power03}.  Fig.~\ref{fig:cam_planck} and
\citet{Ludlow13} include haloes with as few  as $5000$
particles. Since these are cosmological volumes, the samples are
dominated by haloes within an order of magnitude of the limit. The
mean number of particles per halo for the simulations used in
Fig.~\ref{fig:cam_planck} is $N\sim 10000$.  Recall that in
Fig.~\ref{fig:cam} we only include haloes with at least $10^5$
particles.

Since \citet{Ludlow13} use cosmological boxes, lower mass haloes
are poorer resolved. In their Fig.~2 one can clearly see that the
scatter in both $c$ and $\alpha$ is higher for poorer resolved haloes.
By comparing results at a halo mass of $10^{13.6}$M$_{\odot}$ we see
that the haloes from the largest box (and hence lowest resolution)
have higher average $\alpha$ and lower average $c$ than the haloes
from the smallest (and hence higher resolution box). This shows an
anti-correlation between $\alpha$ and $c$, plausibly due to fitting
degeneracies.

By comparing Einasto density profiles with different $\alpha$ and
scale radii, it is easy to see how a fitting degeneracy could
occur. An example is shown in Fig.~\ref{fig:degeneracy1}. For this
example we chose typical values of the scale radius and $\alpha$. All
three profiles are normalized to the scale radius of the input
halo. When $\alpha$ is increased (red line) the profile becomes more
peaked, while lower $\alpha$ (blue line) results in a flatter
profile. The deviations get larger the further away from the scale
radius we go, so with a halo resolved to 1\% of the virial radius
there is a clear distinction between different $\alpha$.  For a less
well resolved halo, e.g., minimum radius of 3\% of the virial radius
one can improve the fit by changing the scale radius (and overall
normalization).

\begin{figure}
\includegraphics[width=0.48\textwidth]{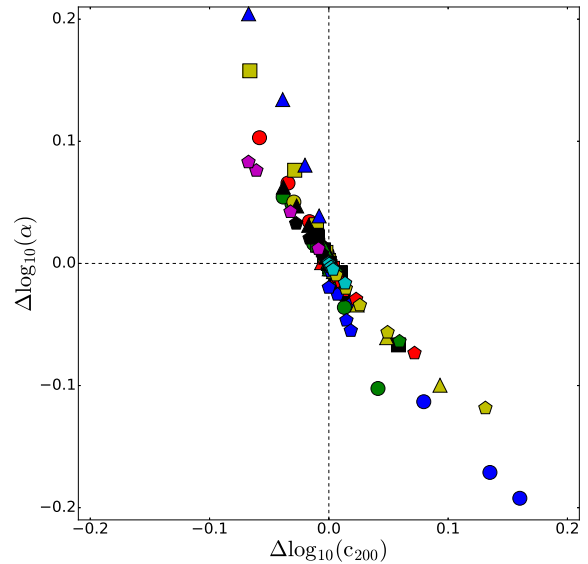}
\caption{Covariance between fitted $\alpha$ and $c$. For each halo
  (different colour/symbol) we vary the minimum radius of the data used
  in the fit and plot the change in $\log\alpha$ and $\log c$ with
  respect to the highest resolution fit.}
\label{fig:degeneracy}
\end{figure}

\begin{figure*}
\centerline{ 
\includegraphics[width=0.48\textwidth]{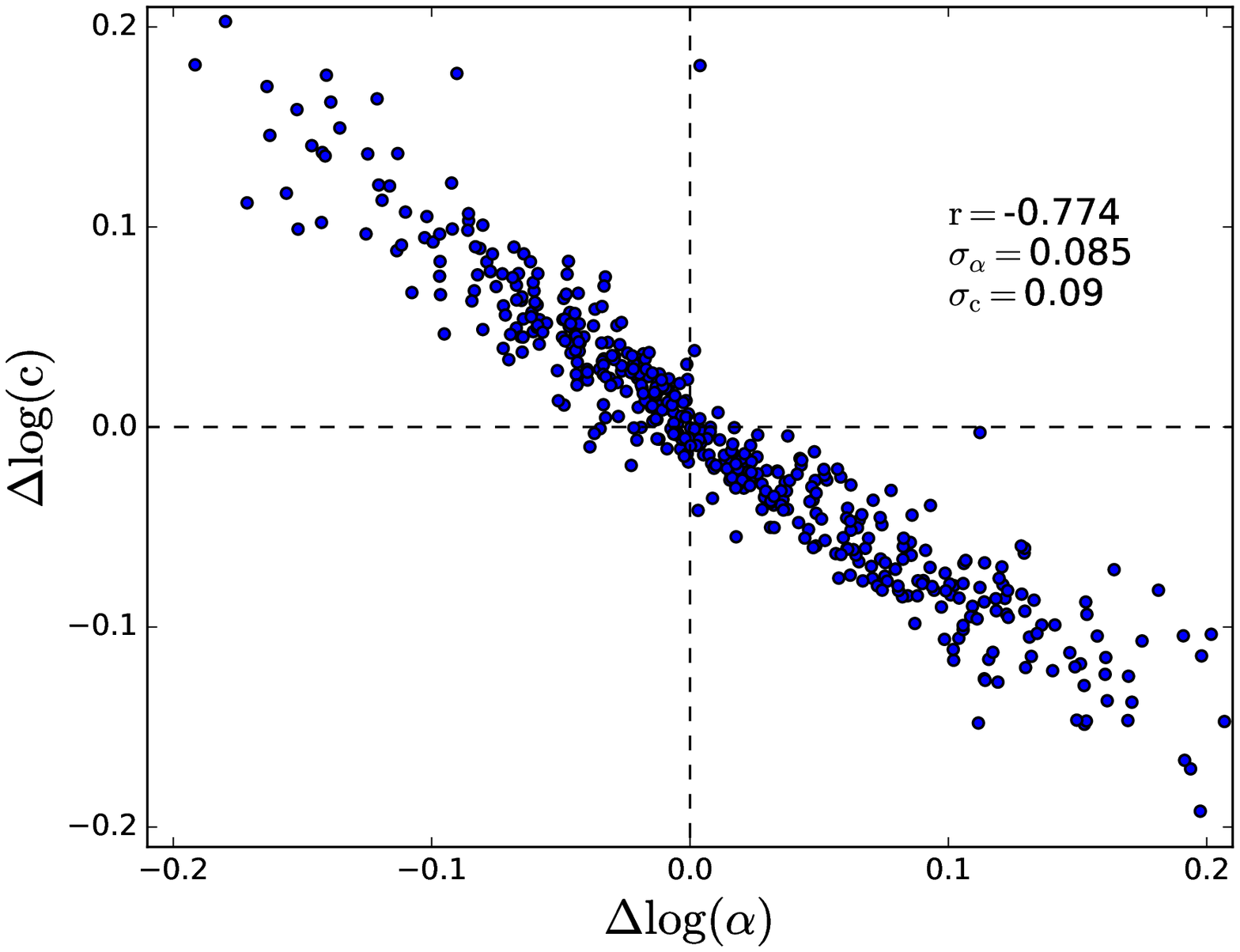}
\includegraphics[width=0.48\textwidth]{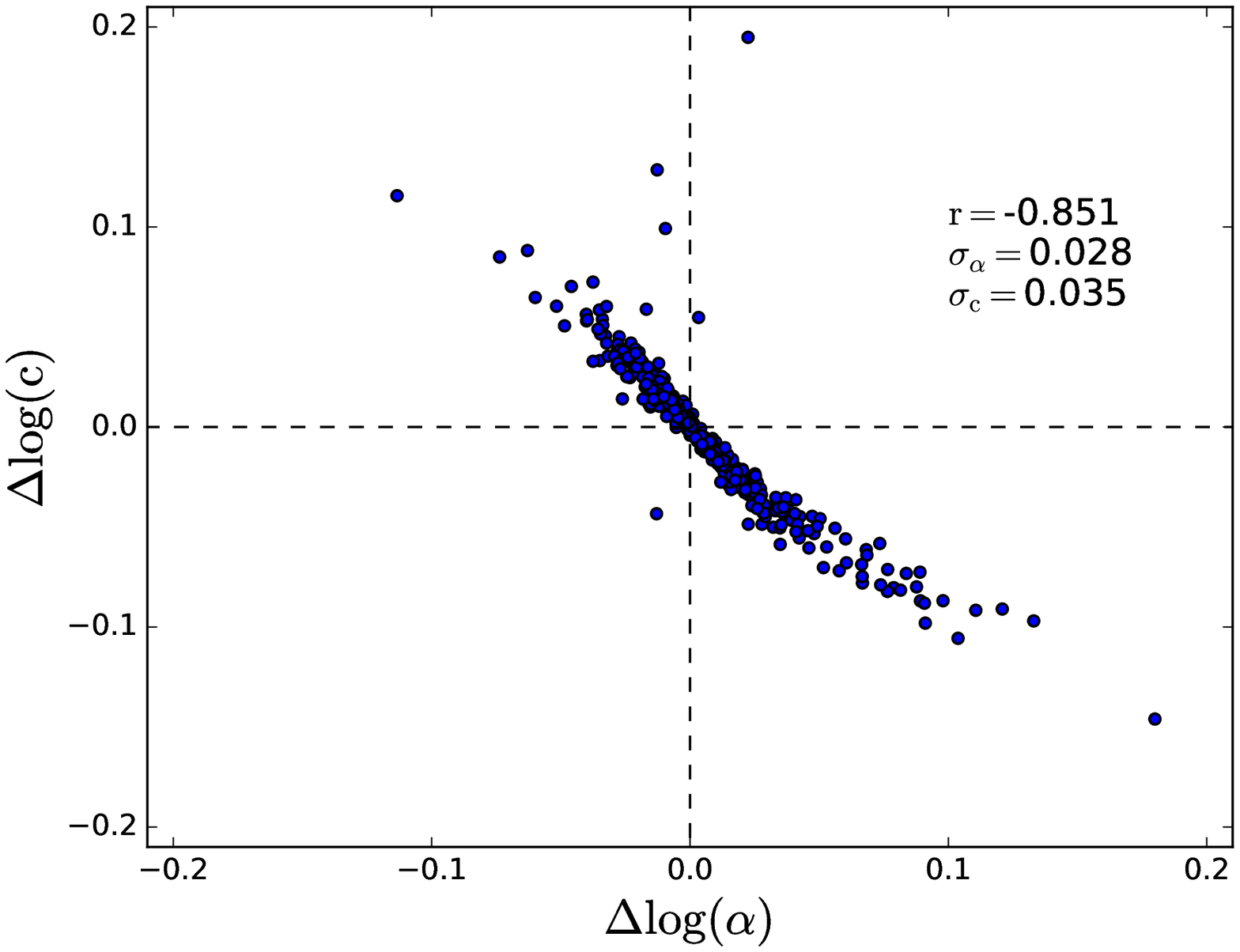}
}
\centerline{
\includegraphics[width=0.48\textwidth]{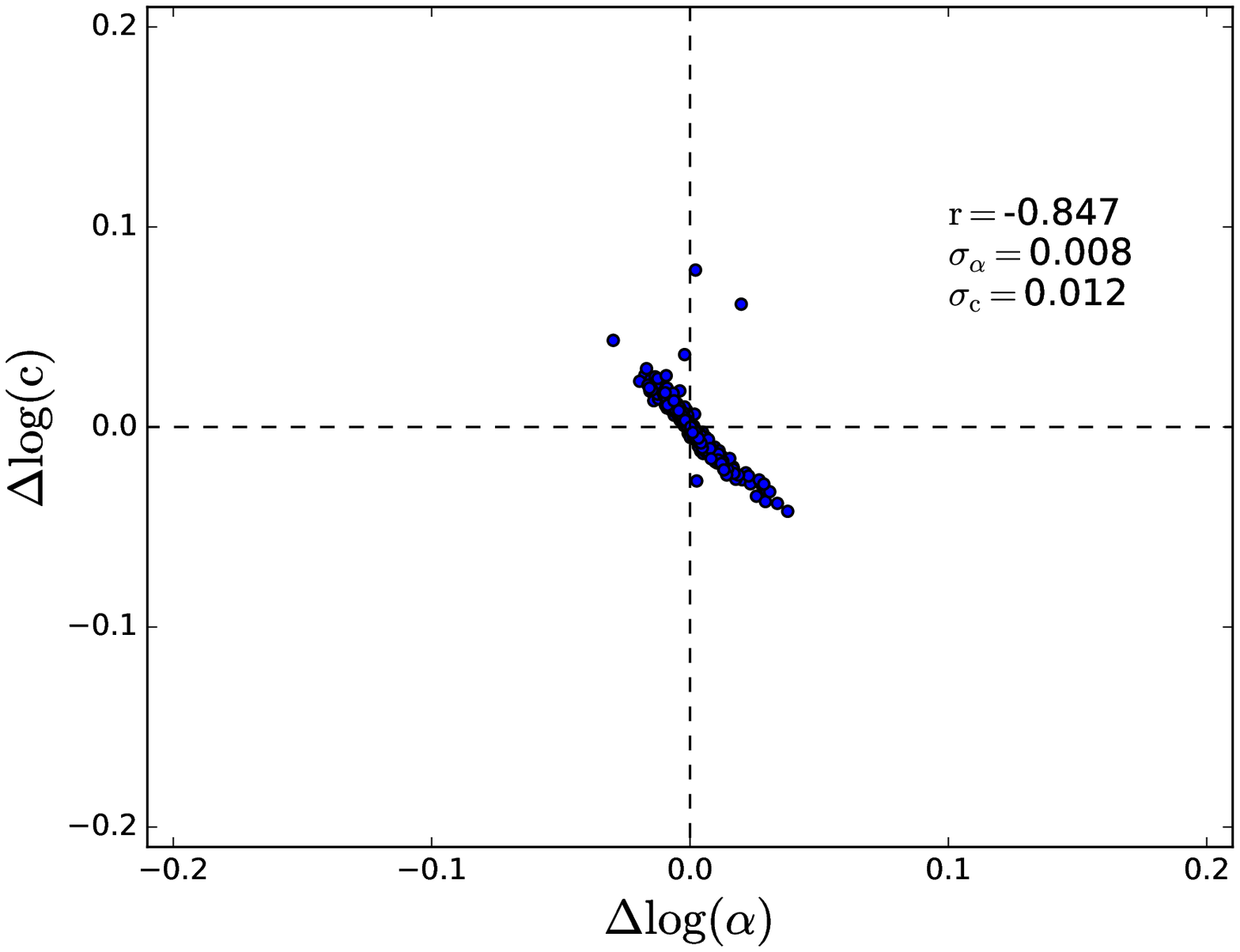}
\includegraphics[width=0.48\textwidth]{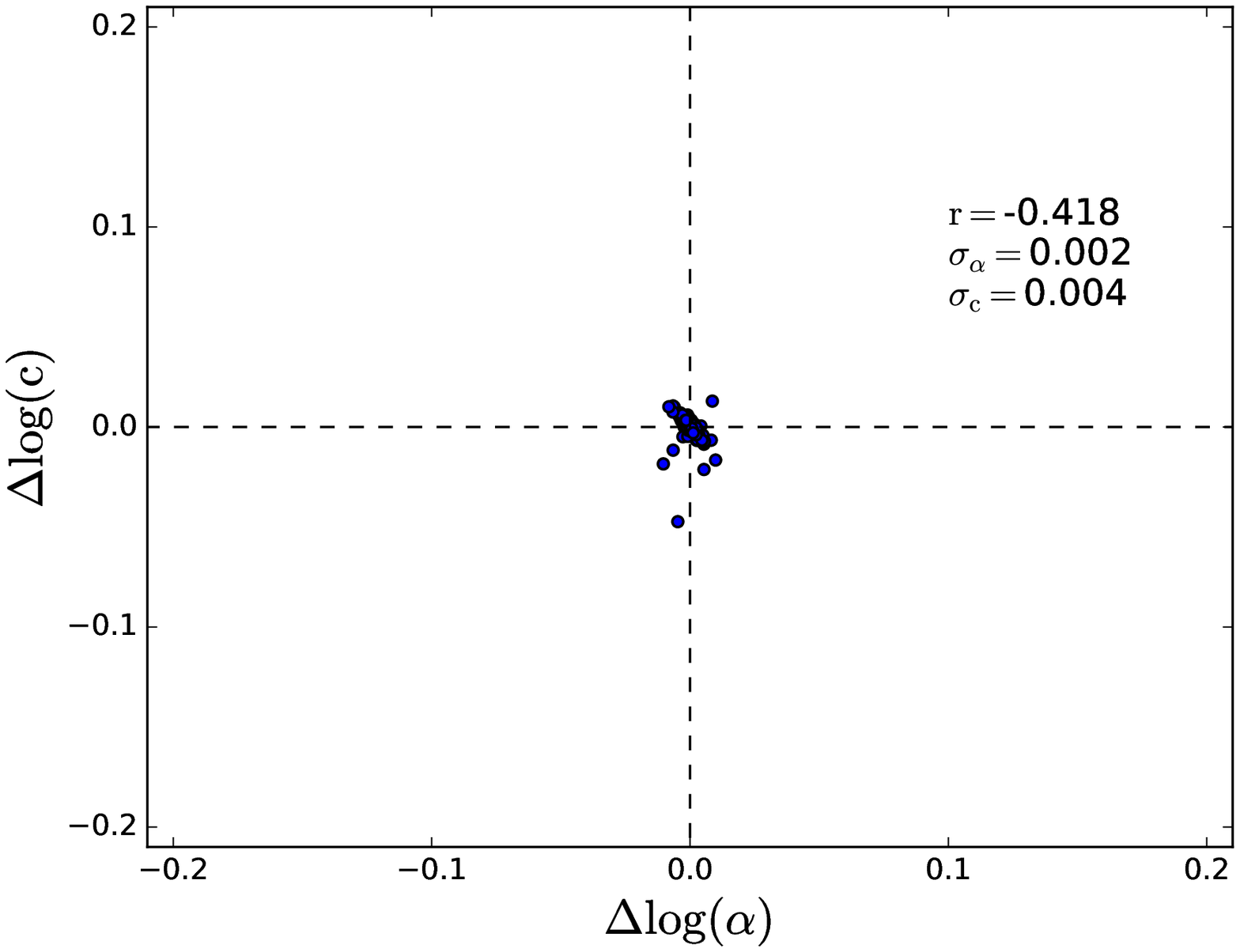}
}
\caption{Covariance between errors on fitted $\alpha$ and $c$ from fits to mock
  Einasto density profiles.  Each panel corresponds to a different
  minimum radius and corresponding number of particles inside the virial
  radius: $N\approx4000$ (top left); $N\approx50000$ (top right);
  $N\approx300000$ (bottom left); $N\approx2$ million (bottom right).
  The numbers in the top right quadrant give the correlation coefficient, $r$,
  the standard deviation of $\Delta \log c$ and $\Delta \log \alpha$.}
\label{fig:mocks}
\end{figure*}

To investigate the role of resolution further we re-fit the Buck \etal
haloes and vary the minimum radius of the density profile to mimic the
effects of lower resolution simulations.  In Fig.~\ref{fig:rmin} we
plot the resulting $\alpha$ and $c$ vs the minimum radius used in the
fit.  We see that the fitted values are quite stable from one
resolution to the next, but there are systematic shifts, so that the
scatter in $c$ and $\alpha$ increases with poorer resolution. In
Fig.~\ref{fig:degeneracy} we plot the change in $\alpha$ (with respect
to the highest resolution fit) vs the change in $c$. Each
colour/symbol corresponds to a different halo. There is a clear
anti-correlation between $\alpha$ and $c$ which confirms our suspicion
that a fitting degeneracy is responsible for the anti-correlation
between $\alpha$ and $c$ found with lower resolution haloes.

Going a step further we create 1000 mock density profiles varying the
halo mass from $10^{9}$ to $10^{13}$. We use   the Einasto function
and draw the parameters ($\alpha, c$) from our fits at redshift $z=0$.
Each profile is sampled from 0.002 to 1.0 $R_{200}$ with 15 equally
log-spaced bins, with noise added to the density profile.
We do a series of fits, successively removing the innermost bin.
In Fig.~\ref{fig:mocks} we plot the error on $\log(c)$ vs the error on
$\log(\alpha)$ for all 1000 mocks. Each panel
corresponds to a different $R_{\rm min}$.  We relate the innermost bin to the
number of particles in the virial radius using the convergence
criteria of \citet{Power03}:
\begin{equation}
  0.6 = \frac{N(R_{\rm conv})}{8\ln N(R_{\rm conv})} \left( \frac{R_{\rm conv}}{R_{200}}\frac{ M(R_{\rm conv})}{M_{200}} \right )^{1/2},
\end{equation}
Given an $R_{\rm min} \equiv R_{\rm conv}$, and a mass profile $M(R)\equiv m_{\rm p}\, N(R)$, we solve for the number of particles within the virial radius.

For $N=2$ million (lower left) the error on $c$ and $\alpha$ is small
$<1\%$.  With poorer resolved haloes the errors on the fitted $c$ and
$\alpha$ increase, to $\sim 7\%$ for $N=50000$ (top right) and $\sim
23\%$ for $N=4000$ (top left).  But more importantly there is a clear
anti-correlation between $c$ and $\alpha$  for $N < 300000$.  For the
results in Fig.2 we use at least $10^5$ particles and typically have
$10^6$ particles, and thus do not expect significant fitting
degeneracies between $c$ and $\alpha$.

In summary we have shown using mock density profiles and re-samples of
our simulated density profiles that poorer resolved haloes suffer a
degeneracy between the fitted $c$ and $\alpha$.  This reconciles our
result of no correlation between $c$ and $\alpha$ (at fixed halo mass)
obtained with a million particles per halo, with the strong
correlation (at fixed halo mass) found by \citet{Ludlow13} obtained
with $\sim 10^4$ particles per halo.  Since there is a strong
correlation between $c$ and mass accretion history, one might be
tempted to conclude that our results imply no correlation between
$\alpha$ and mass accretion history, and thus contradict the result
found by \citet{Ludlow13}. However, the correlation reported by
\citet{Ludlow13} is a weak one, and there is scatter in the relation
between $c$ and mass accretion history. If this scatter depends on
$\alpha$, then it could reconcile our apparently contradictory
results.

\begin{figure*}
  \centerline{
  \includegraphics[width=0.48\textwidth]{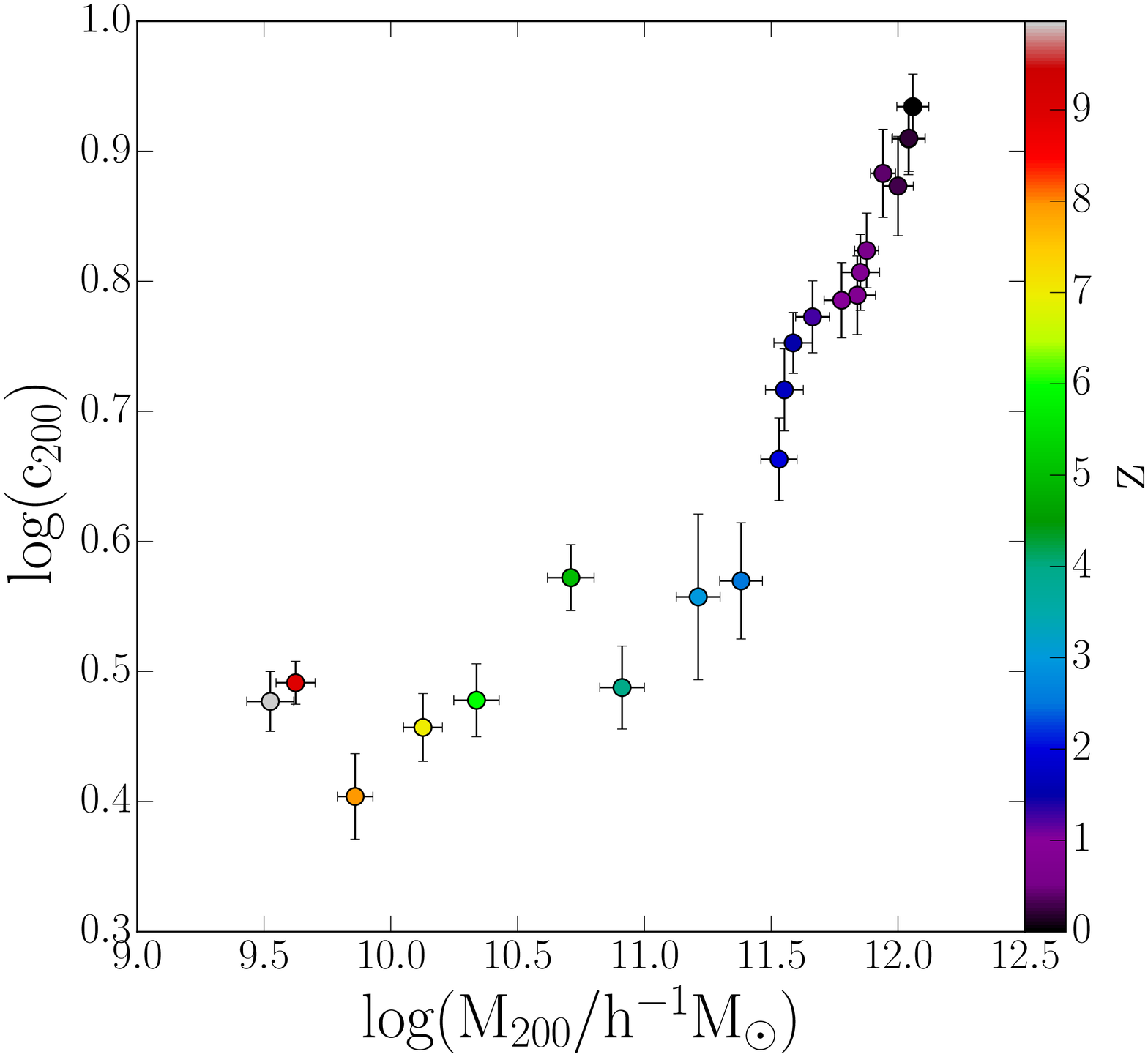}
  \includegraphics[width=0.48\textwidth]{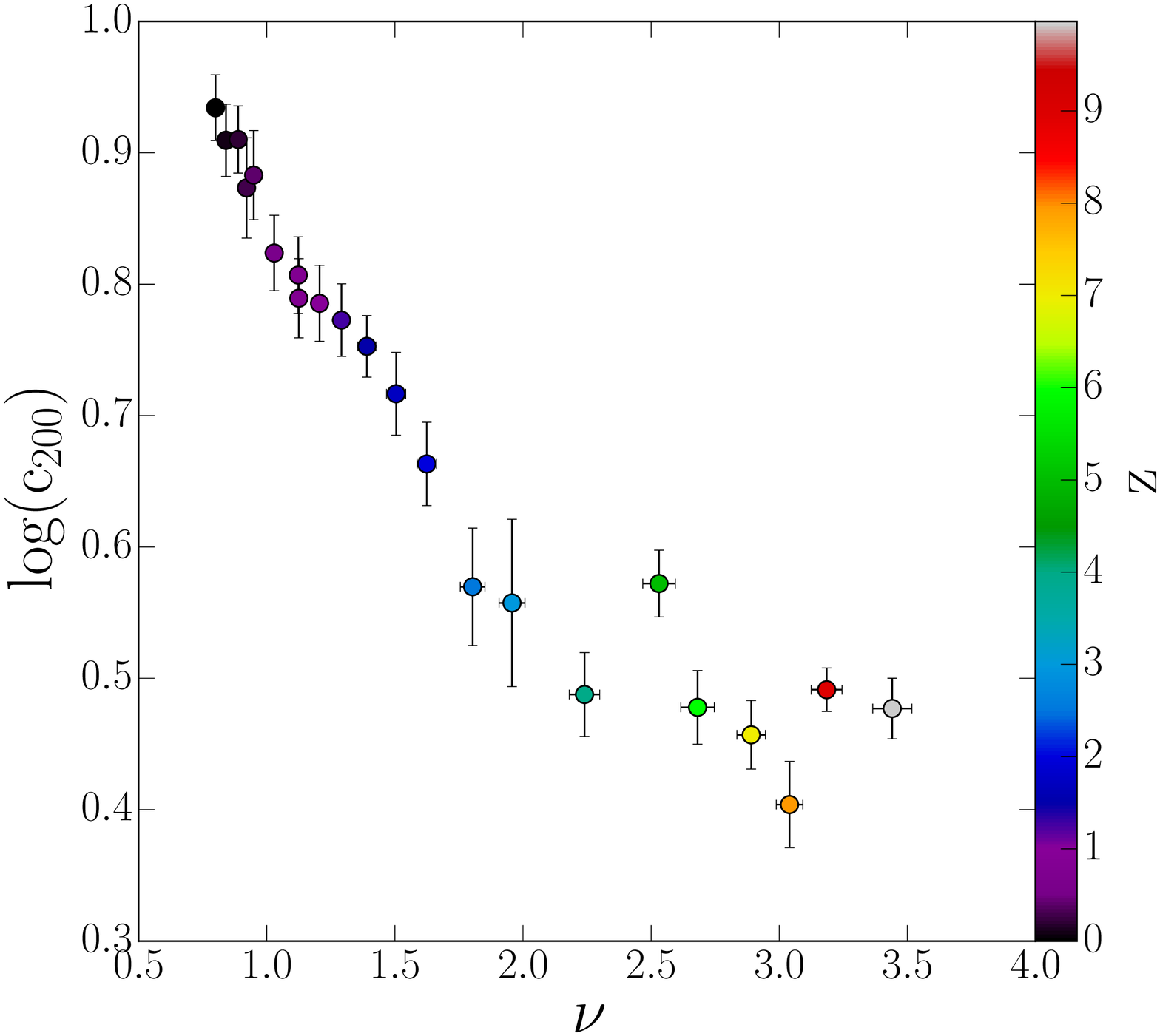}
  }
  \centerline{
  \includegraphics[width=0.48\textwidth]{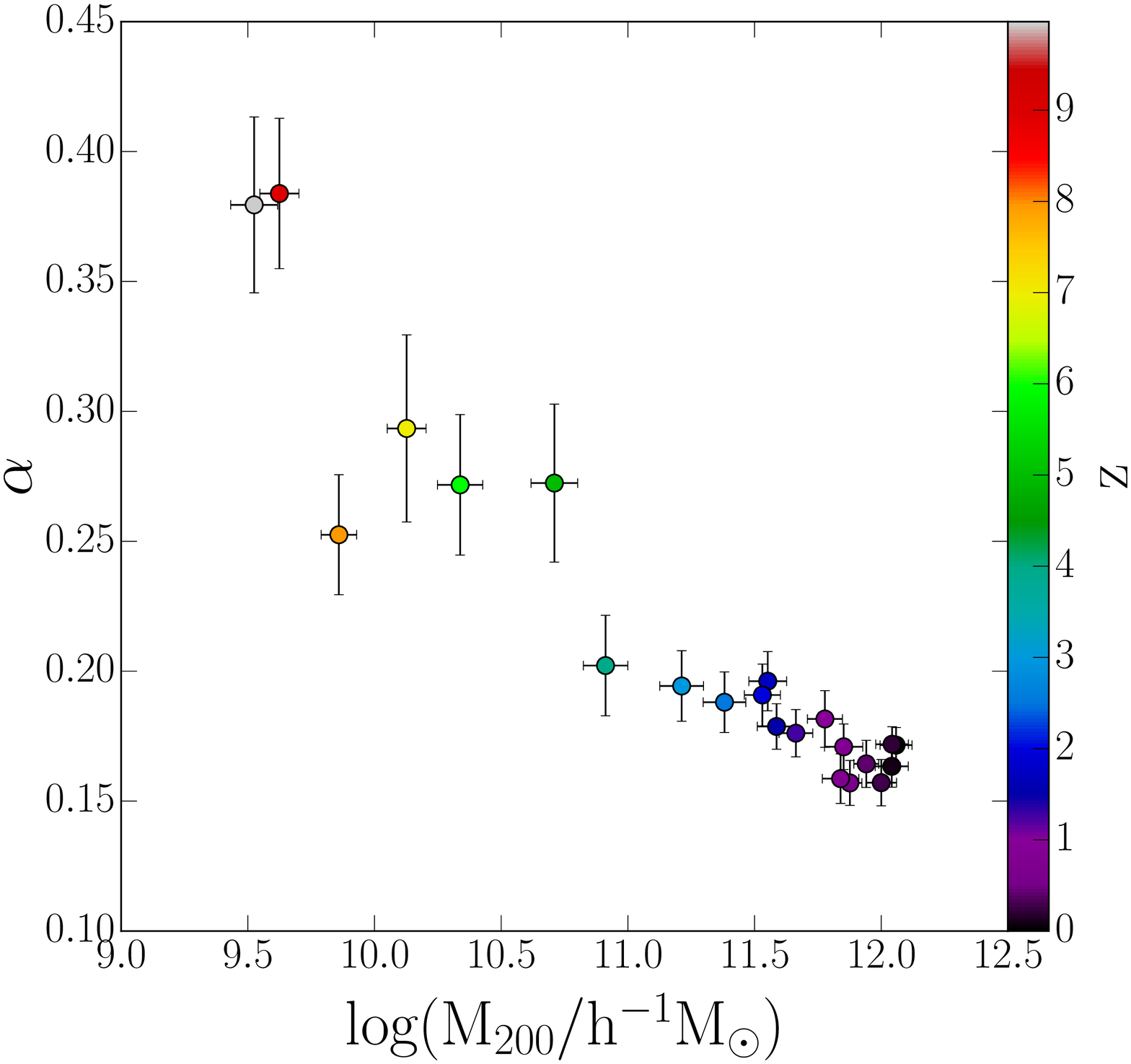}
  \includegraphics[width=0.48\textwidth]{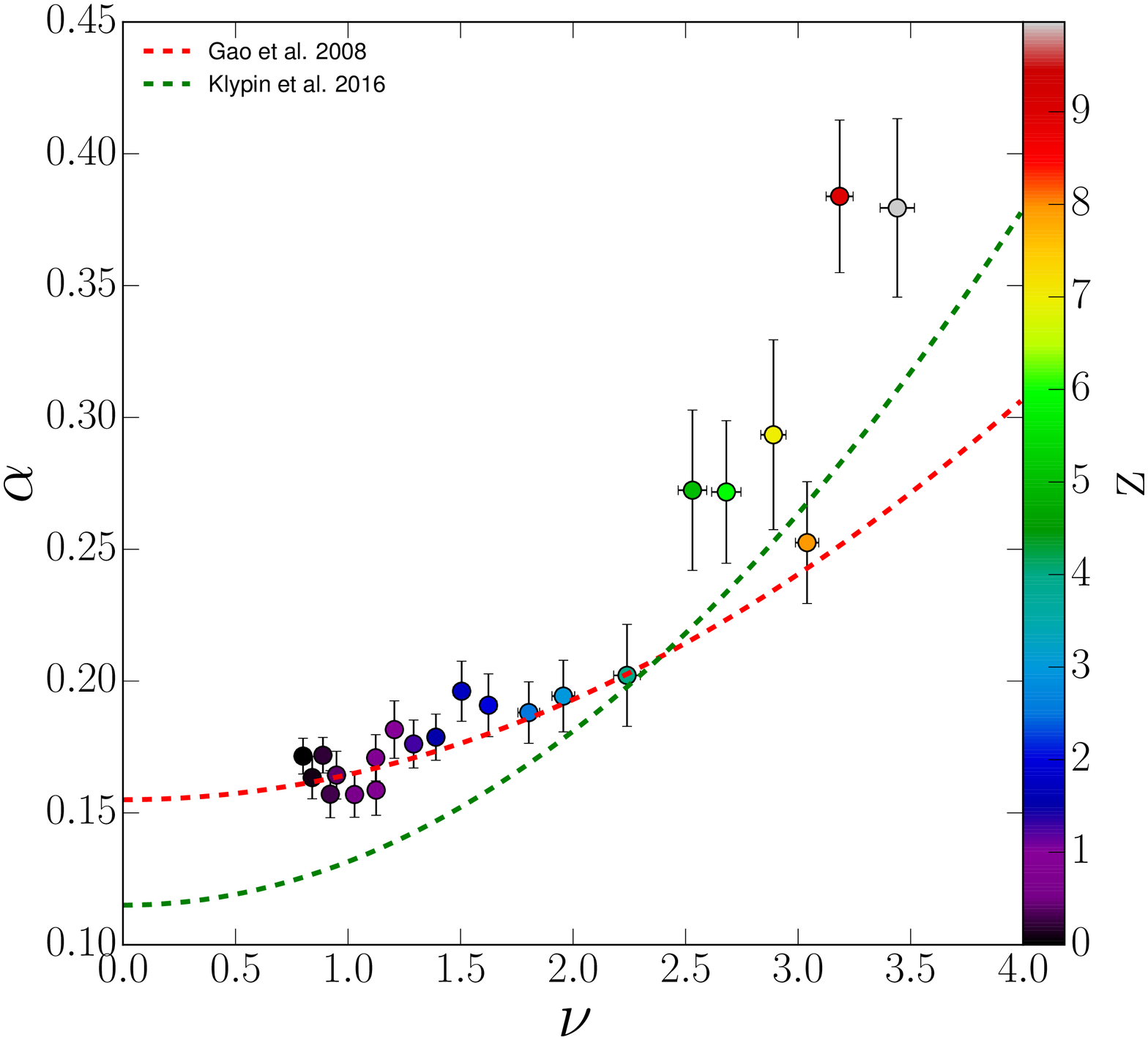}
  }
  \centerline{
  \includegraphics[width=0.48\textwidth]{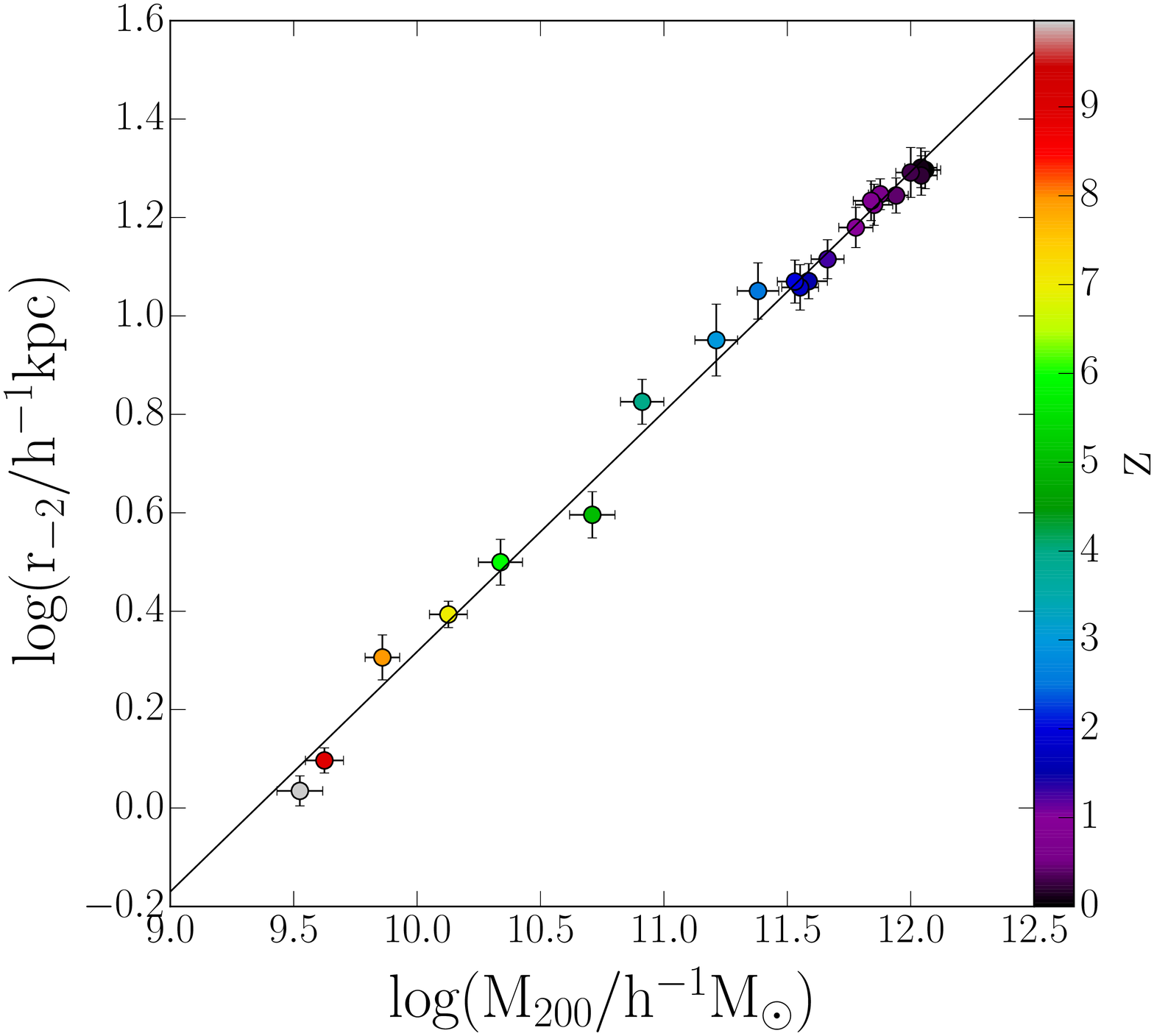}  
  \includegraphics[width=0.48\textwidth]{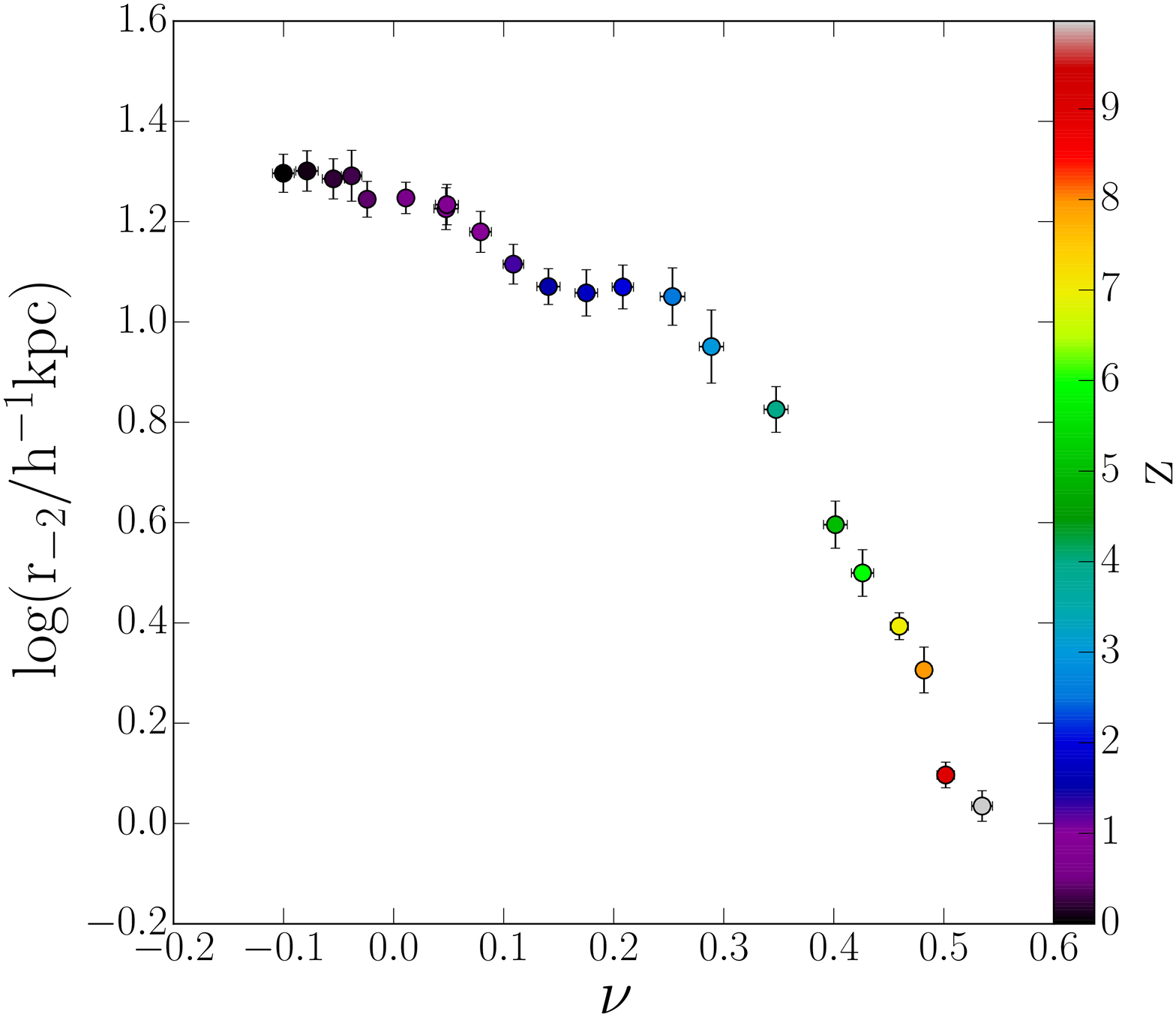}
  } 
\caption{Evolution of the concentration, $c$, Einasto shape parameter,
  $\alpha$, and scale radius, $r_{-2}$, vs halo mass, $M_{200}$ (left), and peak
  parameter, $\nu$ (right).  At each redshift, $z$ (indicated by the colour),
  the median value for the most massive haloes in the Buck \etal
  simulations are plotted. The error bar corresponds to the error on
  the median. For the lower left plot, a power-law was fitted to the
  points (see Equation 3).}
\label{fig:cam2}
\end{figure*}

\begin{figure*}
  \centerline{
  \includegraphics[width=0.5\textwidth]{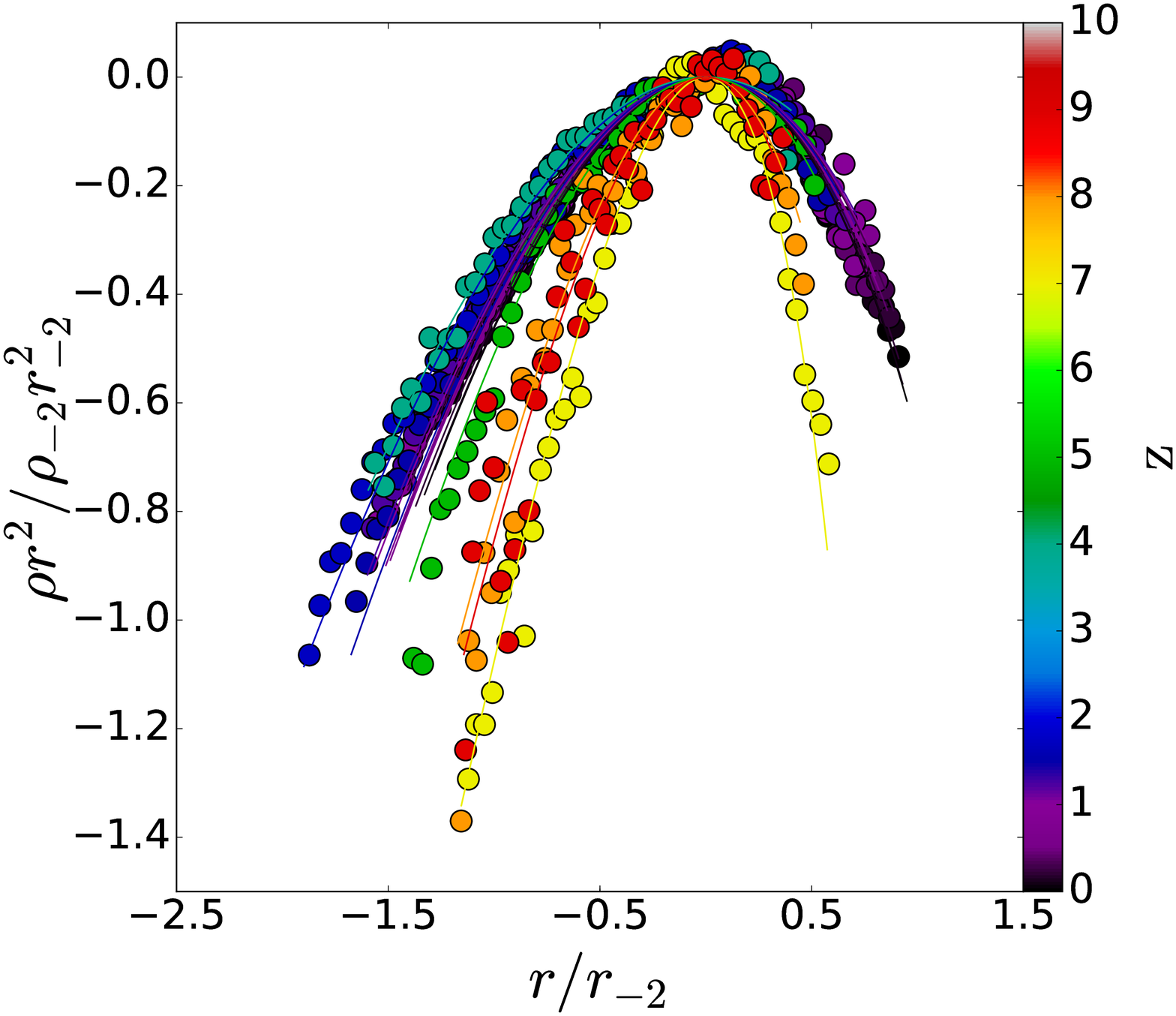}
  \includegraphics[width=0.5\textwidth]{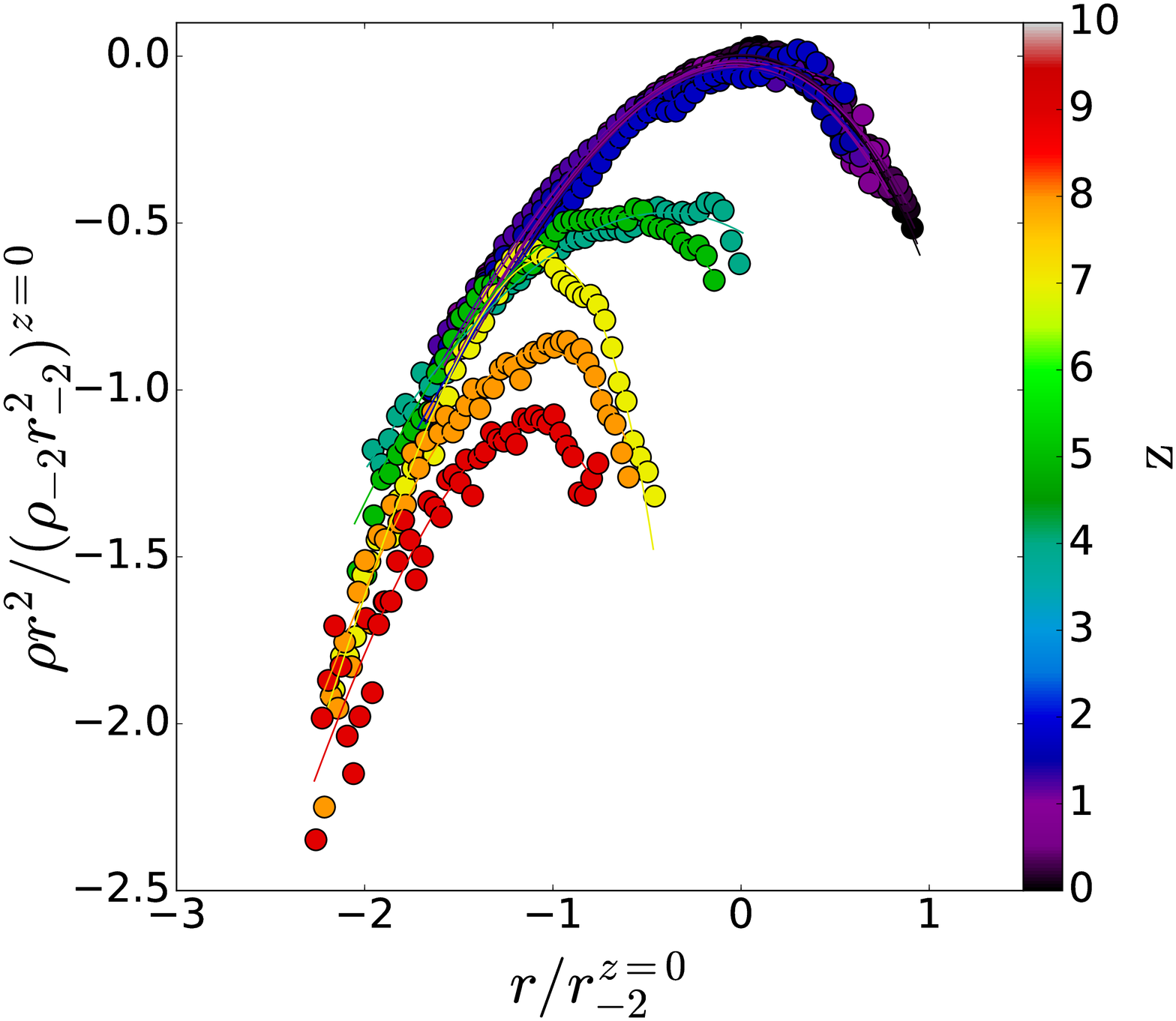}}   
\caption{Evolution of the density profile (coloured points) and Einasto
  fits (solid lines) at different redshifts, for a $10^{12} M_{\odot}$
  halo. For the left panel, the data and fits were normalized to the
  values of $r_{-2}$ and $\rho_{-2}\,r_{-2}^2$ at each individual
  redshift, while for the right panel, the  normalization is relative
  to the values of $r_{-2}$ and $\rho_{-2}\,r_{-2}^2$ at redshift
  $z=0$.}
\label{fig:profiles}
\end{figure*}

\section{Evolution of Most Massive Progenitors}
\label{sec:evolution}

We have shown that there is no significant correlation between
$\alpha$ and $c$ at a given redshift for a sample of well
  resolved haloes.  We now study the evolution of the most massive
(and hence well resolved haloes) in the Buck \etal sample. At each
redshift we have a sample of about 20 haloes.  In order to probe halo
masses below $\sim 10^{10}\msun$ and redshift $z> 7$ we relax the
criteria to include halos with at least $10^4$ particles. The fits at
the highest redshifts will thus be individually less reliable as
those at lower redshifts, nevertheless, the poorer resolved haloes
continue the trends set by the better resolved haloes.  At each
redshift we calculate the median concentration, shape parameter, and
halo mass.  The left-hand panels in Fig.~\ref{fig:cam2} show the
evolution of the $c$ vs mass, $\alpha$ vs mass, and $r_{-2}$ vs mass
relations.

At early times ($z\sim 10$) haloes have low concentration, $c\simeq
3$, small scale radii, $r_{-2}\sim 1 h^{-1}$kpc and high shape
parameter $\alpha \simeq 0.4$. As the halo grows in mass the scale
radius and concentration increases while the shape parameter
decreases relatively smoothly. Thus for the most massive progenitor of
a given halo,  the evolution of $c$ and $\alpha$ are
anti-correlated. Interestingly there is a power-law relation between
the  scale radius and halo mass with a slope of roughly $1/2$:
\begin{equation}
  \log_{10}\left(\frac{r_{-2}}{[h^{-1}{\rm kpc}]}\right)
  =0.488\log_{10}\left(\frac{M_{200}}{[10^{12}h^{-1}\msun]}\right) +1.30.
\end{equation}

In the right-hand panels we replace halo mass with the dimensionless
peak height parameter, $\nu$ (see Eq.~\ref{eq:peakheight}). The peak
height of the most massive halo increases as we go back in time, so
the plots with $\nu$ are almost a mirror reflection of the plots with
mass.  The middle panel shows the well known correlation between
$\alpha$ and $\nu$. The dashed and dotted lines show fitting formula
from \citet{Gao08} and \citet{Klypin16}, respectively which our
simulations are broadly consistent with.

How does the density profile of the haloes evolve?  The scale radius
corresponds to the peak of the $r^2 \rho$ vs radius plot, while the
shape parameter is sensitive to the density profile both at larger and
smaller radii than $r_{-2}$.  In Fig.~\ref{fig:profiles} we show the
evolution of the density profile of a single halo.  The left panel
shows profiles that have been normalized to the scale radius and scale
density at each redshift. We clearly see higher redshift haloes have
density profiles more concentrated (higher $\alpha$) around the scale
radius. Relative to the scale radius, mass is being added at both
small and large radii with time.
In the right panel the profiles are normalized to the scale radius and
density at redshift $z=0$. This shows that after the initial growth at
$z > 7$ the evolution is primarily due to mass being added at
physically large radii.  By $z\sim3$ the density profile is largely in
place, and while the virial radius grows by pseudo evolution of the
background density, the $\alpha$ and $r_{-2}$ appear to continuously
evolve.

\section{Discussion}
\label{sec:disc}
As discussed in the introduction it has been   suggested that the
slope of the power-spectrum of density fluctuations determines the
Einasto $\alpha$ parameter \citep{Cen14, Nipoti15, Ludlow17}. If this
were the sole explanation, then we would expect higher mass and $\nu$
haloes to have lower $\alpha$ in $\Lambda$CDM simulations. The fact the
opposite is seen means there must be another explanation.

In our simulations we find the most massive progenitors of Milky Way
mass haloes have high $\alpha$, which subsequently decreases with
increasing time. This is reminiscent of the violent relaxation process
\citep{Lynden-Bell67} of dissipationless major mergers which tends to
decrease $\alpha$ towards, but does not reach, the isothermal value
($\alpha=0$). We thus speculate that it is the evolutionary state
of the halo that determines $\alpha$. 

The connection with the power-spectrum slope is due to the fact that
in a given amount of time, major mergers have a much larger relaxation
effect than an equivalent sum of minor mergers \citep{Hilz12}.  This
is related to the fact the dynamical friction time scale, $t_{\rm df}
\propto (M/m) / \ln(M/m)$, where $M/m$ is the mass ratio, is shorter
for major mergers. Thus the haloes formed from flat spectra will
experience more major mergers and are expected to be more dynamically
evolved than haloes of the same mass that form from steep spectra via
minor mergers.

We expect the evolutionary trend of $\alpha$ decreasing is also true
for higher and lower mass haloes (because of the $\alpha$ vs $\nu$
relation appears independent of redshift). Nevertheless, it would be
worth verifying this with high-resolution simulations.

\section{Summary}
\label{sec:conc}
We use high resolution (up to 10 million particles per halo) N-body
zoom-in simulations to study the structure of dwarf to Milky Way mass
dark matter haloes over cosmic time. We fit our haloes with the three
parameter Einasto function, specified by the halo mass, concentration,
$c$, and profile shape, $\alpha$.  We summarize our results as
follows:

\begin{itemize}
\item At a given redshift, $\alpha$ is independent of $c$
  (Fig.~\ref{fig:cam}). This means that (at least) three parameters
  are needed to accurately describe the structure of $\Lambda$CDM
  haloes.

\item We trace previous reports \citep{Ludlow13} of a strong
  anti-correlation between $\alpha$ and $c$ to a fitting degeneracy in
  lower resolution ($\sim 10^4$ particles per halo) simulations
  (Figs.~\ref{fig:degeneracy} \& \ref{fig:mocks}). 

\item However, for the most massive progenitors of individual haloes
  the evolution in $\alpha$ and $c$ is anti-correlated (Fig.~\ref{fig:cam2}).
  At high
  redshift ($z\sim 7$) $\alpha\sim 0.3$ and $c\sim 3$. As the halo
  evolves $\alpha$ decreases while $c$ increases.
  
\item The increase in concentration is due to an increase in virial
  radius, which more than compensates for the increase in scale
  radius, $r_{-2}$  with time (Fig.~\ref{fig:cam2}). 

\item The evolution of halo structure ($\alpha$ and $r_{-2}$) is
  primarily due to accretion of  mass at large radii
  (Fig.~\ref{fig:profiles}).
\end{itemize}

We speculate that the evolution of $\alpha$ is related to the
relaxation state of the halo. Younger, less dynamically relaxed haloes
have higher $\alpha$ closer to a spherical top hat, while older, more
dynamically relaxed haloes have lower $\alpha$ closer to an
isothermal. 
Studying the dynamics of the halo particles would be a
way to verify this.

\section*{Acknowledgments}

We thank the anonymous referee whose comments helped to improve the clarity of the paper.
This research was carried out on the High Performance Computing
resources at New York University Abu Dhabi; on the  {\sc theo} cluster
of the Max-Planck-Institut f\"ur Astronomie and on the {\sc hydra}
clusters at the Rechenzentrum in Garching. We greatly appreciate the
contributions of all these computing allocations.
%


\end{document}